\documentclass[aps,prl,twocolumn,showpacs,superscriptaddress,groupedaddress]{revtex4}  
\usepackage{graphicx}  
\usepackage{dcolumn}   
\usepackage{bm}        
\usepackage{amssymb}   
\usepackage{amsmath}
\usepackage{xcolor}
\usepackage{amsthm}
\usepackage{enumitem}
\newtheorem{theorem}{Theorem}

\begin{document}


\title{Maximum Entropy Principle in statistical inference:\\ case for non-Shannonian entropies}

\author{Petr Jizba}
\email{p.jizba@fjfi.cvut.cz}
\affiliation{FNSPE,
Czech Technical University in Prague, B\v{r}ehov\'{a} 7, 115 19, Prague, Czech Republic}

\author{Jan Korbel}
\email{jan.korbel@meduniwien.ac.at}
\affiliation{Section for Science of Complex Systems, Medical University of Vienna, Spitalgasse 23, 1090 Vienna, Austria}
\affiliation{Complexity Science Hub Vienna, Josefst\"{a}dter Strasse 39, 1080 Vienna, Austria}
\affiliation{FNSPE,
Czech Technical University in Prague, B\v{r}ehov\'{a} 7, 115 19, Prague, Czech Republic}

\begin{abstract}

In this Letter we show that the Shore--Johnson axioms for Maximum Entropy Principle in statistical estimation theory account
for a considerably wider class of entropic functional than previously thought. Apart from a
formal side of the proof where a one-parameter class of admissible entropies is identified, we substantiate our point by analyzing the effect of weak correlations and by discussing two pertinent examples:
$2$-qubit quantum system and transverse-momentum behavior of hadrons in high-energy  proton-proton  collisions.
\end{abstract}
%

\vspace{-1mm}
\maketitle
%
%

\vspace{-3mm}
The concept of entropy indisputably plays a pivotal role in modern
physics~\cite{Tsallis-book,beck}, statistics~\cite{millar,shore,shoreII}  and information theory~\cite{jaynes3,petz}.
In each of these fields  the entropy paradigm has been formulated independently
and with different
applications in mind. While
in physics the entropy quantifies the number of distinct microstates compatible
with a given macrostate,  in statistics it corresponds to the inference functional for an updating procedure,
and in information theory it determines a limit on
the shortest attainable encoding scheme.

However, recent developments in quantum theory~\cite{biro,biro_II} and complex dynamical systems in particular~\cite{thurner,thurner-book,thurner_II,thurner_IIa}
have brought about the need for a further extension of the concept of entropy beyond conventional Shannon--Gibbs type of entropies.
Consequently, numerous generalizations proliferate in the current literature ranging from additive entropies of R\'{e}nyi~\cite{jizba3} and Burg~\cite{burg} through rich class of non-additive entropies~\cite{tsallis2,havrda,sharma,frank,arimitsu,korbel} to more exotic types of entropies~\cite{segal}. Concomitantly
with this, efforts are under way to classify all feasible entropic functionals according to their group properties~\cite{tempesta},
generalized additivity rules~\cite{ilic} or asymptotic scaling~\cite{hanel,thurner_IIa}.

Regardless of a particular generalization, the key usage of entropy is in statistical estimation theory, which
in turn crucially hinges on the Maximum entropy (MaxEnt) principle (MEP) and its various reincarnations
(e.g., the maximum likelihood estimate, 
principle of minimum cross-entropy,
minimum Akaike information criterion, etc.). The MEP
can be formulated as follows~\cite{jaynes1,jaynes2,jaynes3}:

\begin{theorem}[{\bf MEP}]
Given the set of constraints $C= \{{I}_k\}_{k=1}^{\nu}$, the best estimate of the underlying (i.e., true) probability
distribution $P = \{ p_i\}_{i=1}^n$ is the one that maximizes
the entropy functional $S[P]$ subject to the constraints, i.e., it maximizes the Lagrange functional
\begin{equation}
S[P]  \ - \ \sum_{k=1}^{\nu} \lambda_k I_k\, .
\end{equation}
\end{theorem}
In the case of inductive inference
the constraints, or prior information,
are given in terms of linear expectation values, i.e., the constraints considered are of the form
\begin{equation}
I_k \ \equiv \ \langle \mathcal{I}_k \rangle \ = \  \sum_i \mathcal{I}_{k, i}\, p_{i} \, ,
\label{constraint.2}
\end{equation}
where $\{\mathcal{I}_{k, i}\}$ are possible realizations (alphabet) of the observable $\mathcal{I}_k$.
To avoid Czisz\'{a}r-type paradoxes, it must be assumed that $C$ singles out a closed (in $\ell^1$-norm) convex
subset of probability distributions in which the true distribution falls~\cite{Cziszar}.
Other types of constraints, such as escort means, quasi-linear means or
non-inductive
prior information, such as Lipshitz--H\"{o}lder exponent of probability distributions or Hausdorff dimension of the state space
are not considered at this stage.

The heuristic justification behind the MEP is typically twofold: first, maximizing entropy minimizes the
amount of prior information built into the distribution (i.e. MaxEnt distribution is maximally noncommittal
with regard to missing information);
second, many physical systems tend to move towards (or concentrate extremely close to) MaxEnt configurations
over time~\cite{Tsallis-book,jaynes1,beck,thurner-book}.

MEP was pioneered by Jaynes who first employed Shannon's entropy in the framework of equilibrium statistical physics~\cite{jaynes1,jaynes2}.
On a formal level the passage from Shannon's information theory to statistical thermodynamics is remarkably simple, namely a MaxEnt probability
distribution subject to constraints on average energy, or average energy and number of particles yield the usual Gibbs' canonical or grand-canonical
distributions, respectively. In classical MEP the MaxEnt distributions are always of an exponential form when constrains are phrased in terms of a
finite number of moments (situation typical in practice).
Applicability of MEP is, however, much wider.
Aside from statistical thermodynamics, MEP has now become a powerful tool in non-equilibrium statistical physics~\cite{kapur} and
is equally useful in such areas as astronomy, geophysics, biology, medical diagnosis and economics~\cite{thurner-book,kapur}.

As successful as Shannon's information theory has been, it is clear by now that it is capable of dealing with only a limited class of systems.
In fact, only recently it has become apparent that there are many situations of practical interest requiring
more ``exotic'' statistics which does not conform with the canonical prescription
of the classical MaxEnt (known as Boltzmann--Gibbs statistics)~\cite{thurner-book}.
On the other hand, it cannot be denied that MaxEnt approach deals with statistical systems in a way that is
methodically appealing, physically plausible and intrinsically nonspeculative (MaxEnt invokes no
hypotheses beyond
the evidence that is in the available data).
One might be thus tempted to extend MEP also on other entropy functionals particularly when the ensuing MaxEnt
distributions differ from Boltzmann--Gibbs ones in some desirable way (e.g. in particular types of heavy tails).
Entropy functionals in question
should not be, however, arbitrary but they ought to satisfy some ``reasonable'' properties.
From the point of information theory, these properties are typified by coding theorems~\cite{Campbell1965,Nath} or axiomatic rules (\`{a} la Shannon--Kchinchine type of axioms~\cite{jizba3,abe}).
Recently, however, doubts have been raised about feasibility of this program.
Arguments involved primarily rest on  Shore--Johnson (SJ) axioms of statistical estimation theory.

{\em {Shore--Johnson axioms.}}~---~
From the point of statistics, MEP is an estimation method, approximating probability distribution from the limited prior information. As such, it should obey some consistency rules.
SJ introduced a set of axioms, which ensure that the MEP
estimation procedure is consistent with desired properties of
inference methods. These axioms are~\cite{shore}:\\[-6mm]
\begin{enumerate}
\item \emph{Uniqueness:} The result should be unique.\\[-6.5mm]
\item \emph{Permutation invariance:} The permutation of states should not matter.\\[-6.5mm]
\item \emph{Subset independence:} It should not matter whether one treats disjoint
subsets of system states in terms of separate conditional distributions or in
terms of the full distribution.\\[-6.5mm]
\item \emph{System independence:} It should not matter whether one accounts for independent
constraints related to independent systems separately in terms of
marginal distributions or in terms of full-system constraints and joint distribution.\\[-6.5mm]
\item \emph{Maximality:} In absence of any prior information, the uniform distribution should be the solution.\\[-6mm]
\end{enumerate}
To keep our discussion as simple as possible we focus on discrete probabilities only.
Let us note, that for continuous probability distributions, the entropy (or better its continuous counterpart --- differential entropy) is not a coordinate-invariant and one must consider the Maximum Relative Entropy principle instead of MEP. The  generalization of the SJ axioms for continuous distributions was discussed, e.g., in Refs.~\cite{shore,uffink}, and results obtained here are (with minor adjustments) valid also in continuous-state spaces.

In recent years, there has been much discussion of the consistency of MEP for generalized, i.e., non-Shannonian entropies. A typical claim has been that the SJ axioms
preclude the use of MEP for generalized entropies, since they introduce an extra bias in the estimation of the ensuing MaxEnt distributions~\cite{presse,tsallis,presse2}.
If this was true then in some important cases, such as in the R\'{e}nyi entropy-based signal processing and pattern recognition, there would be important new
corrections or inconsistencies to some existing analyzes. Here we show that the SJ axioms as they stand {\em certainly} allow
for a wider class of entropic functional than just Shannon's entropy (SE).
Central to this is the following theorem due to Uffink~\cite{uffink}.
\begin{theorem}[{\bf Uffink Theorem}]
MEP satisfies Shore--Johnson consistency axioms if and only if the following prescription holds: Maximize $\mathcal{U}_q(P)$ under the set of constraints $C= \{{I}_k\}$, where $\mathcal{U}_q(P) = \left(\sum_{i=1}^n p_i^q\right)^{1/(1-q)}$
for any $q > 0$, modulo equivalency condition.
\end{theorem}
The {\em equivalency condition}
means that all functionals $f(\mathcal{U}_q(P))$  for strictly increasing functions
$f$ are equivalent ($\sim$) in the sense that they provide the same MaxEnt distribution~\cite{uffink}.
A simple variant of the proof together with related discussion is provided in Supplemental Material~\cite{note}.
Here we list some pertinent results:

(a) From axioms 1-3 alone follows that the entropy is equivalent to the sum-form
functional:
\begin{equation}
\mathcal{U}(P) \ = \ \sum_{i=1}^{n} g(p_i) \ \sim  \ f\left(\sum_{i=1}^{n} g(p_i)\right) .
\end{equation}
Axioms 1-3 alone thus rule out a wide class of existent  entropies. Examples include: $(a,\lambda)$-escort entropies~\cite{bercher}; \mbox{$S_{a,\lambda}(P) = 1/(\lambda-a) \left[\left(\sum_i p_i^a\right)^{\lambda} \left(\sum_i p_i^\lambda\right)^{-a}-1 \right]$}
or JA hybrid entropy~\cite{arimitsu,korbel}; $\mathcal{D}_q(P) = \ln_q \exp [-\sum_i P_i(q) \ln p_i]$, where $\ln_q x = (x^{1-q}-1)/(1-q)$ is the $q$-logarithm and $P_i(q) = p_i^q/\sum_j p_j^q$ is the escort distribution~\cite{beck}.
(b) Axiom 4 ensures that any entropy functional consistent with SJ axioms should be {\em equivalent} to $\sum_i p_i^q$.
There is a number of entropic functionals that do not conform to this form, examples include: $(c,d)$-entropy~\cite{hanel,thurner}; \mbox{$S_{c,d}(P) = \sum_i \Gamma(1+d,1-c\log p_i)$} or the Kaniadakis
entropy~\cite{Kaniadakis}; $S_{\kappa}(P) = \frac{1}{2 \kappa} \sum_i \left(p_i^{1+\kappa} + p_i^{1-\kappa}\right)$.
(c) Axioms 5 implies that inference functional should be of the form:
\begin{equation}
\mathcal{U}_q(P) \ = \ \left(\sum_i p_i^q\right)^{1/(1-q)} \;\;\;\;\mbox{for}   \;\;\;\;\;\;\;\;\; q > 0\, ,
\end{equation}
(modulo equivalency condition). Only for $q > 0$ it is guaranteed that  $\mathcal{U}_q(P)$ is
Schur-concave which is a sufficient property for {\em maximality} axiom~\cite{note}.
For example, Burg entropy~\cite{gorban} $\mathcal{K}(P) = K \sum_i \ln p_i$ provides an example of entropy functional belonging to the class of $\mathcal{U}_q(P)$, but not fulfilling the {\em maximality} axiom. (d)
Shannon's entropy (SE) is a unique candidate for MEP only when an extra desideratum is added to SJ axioms, namely;
\emph{Strong system independence (SSI):} Whenever two subsystems of a system are disjoint, we can treat the subsystems in terms of independent distributions.

So far, the additivity property of the entropy functional was not our concern. Note, however, that functionals $\mathcal{U}_q(P)$ --- known also as R\'{e}nyi entropy powers~\cite{jizba}, obey the multiplicative composition rule $\mathcal{U}_q(A \cup B) = \mathcal{U}_q(A) \mathcal{U}_q(B)$ for independent events. By choosing appropriately $f$, we can construct entropies with various types of composition rules. For instance, for $f(x) = \ln x$, we get a class of additive R\'{e}nyi entropies (including Shannon's one)~\cite{jizba3}, if $f(x) = \ln_\mathcal{Q} x$ is chosen, we obtain $\mathcal{Q}$-additive Sharma--Mittal entropies~\cite{sharma}. For $\mathcal{Q}=q$ we end up with the class of Tsallis entropies~\cite{Tsallis-book}. Consequently, the MEP procedure implied by SJ axioms
does not preclude, per se, any additivity rule as long as the entropy is  $\sim \mathcal{U}_q(P)$.

Despite this, it is asserted in a number of recent works, cf., e.g.,~\cite{presse,tsallis,presse2},  that the only inference functional consistent with the SJ desiderata is SE, i.e., the $q=1$ case.
This was also the original result of SJ. The point of disagreement with these works can be retraced back to the axiom of {\em system independence} and its implementation in the original SJ proof~\cite{shore,shoreII}.
Notably, SJ assumed that because the prior distributions $Q_1$ and $Q_2$ are independent (in MEP they are uniform) and because the data-driven constraints $I_1$ and $I_2$
are independent (i.e., they give no information about any interaction between the two systems), the posterior distribution $P$ must
be written as a product of marginal distributions $U$ and $V$.
However, this  goes well beyond the original SJ axiom 4 
in that the presumed independency of constraints invokes (unwarranted though often reasonable) unique factorization rule for the posterior.
Clearly, having no information about interaction encoded in constraints
(i.e., having independent constraints) is not the same as having no correlations among systems.
Let us now show that there is an implicit assumption about the state-space structure in the SJ proof
yielding the specific factorization rule.

{\em {Factorization rule revisited.}}~---~
Let us now concentrate on the composition rule of MaxEnt distributions for two  systems
described by marginal distributions $U = \{u_i\}_{i=1}^n$ and $V = \{v_j\}_{j=1}^m$ and related constraints
$\sum_{i=1}^n  \mathcal{I}_i u_i = \langle \mathcal{I} \rangle $ and $\sum_{j=1}^m \mathcal{J}_j v_j = \langle \mathcal{J} \rangle$. The MaxEnt distributions $U$ and $V$ are obtained by maximizing  $\mathcal{U}_q(U)$ and $\mathcal{U}_q(V)$, respectively.
Ensuing equations read
\begin{eqnarray}
&&\mbox{\hspace{-5mm}}\frac{q}{1-q} \ \![\mathcal{U}_q(U)]^q u_i^{q-1} \ - \ \alpha_\mathcal{I} \ - \ \beta_\mathcal{I} \mathcal{I}_i \ = \ 0\, , \label{eq:1} \\[1mm]
&&\mbox{\hspace{-5mm}}\frac{q}{1-q} \ \![\mathcal{U}_q(V)]^q v_j^{q-1} \ - \ \alpha_\mathcal{J} \ - \  \beta_\mathcal{J} \mathcal{J}_j \ = \ 0\, . \label{eq:2}
\end{eqnarray}
The solutions can be written as
\begin{eqnarray}
&&\mbox{\hspace{-5mm}}u_i \ = \ [\mathcal{U}_q(U)]^{-1} \left[1 \ - \ (q-1) \frac{\beta_\mathcal{I} \Delta \mathcal{I}_i}{q \mathcal{U}_q(U)} \right]^{1/(q-1)}\, ,\\
&&\mbox{\hspace{-5mm}}v_j \ = \ [\mathcal{U}_q(V)]^{-1} \left[1 \ - \ (q-1) \frac{\beta_\mathcal{J} \Delta \mathcal{J}_j}{q \mathcal{U}_q(V)} \right]^{1/(q-1)}\, ,
\label{Eq.7-8}
\end{eqnarray}
with $\Delta \mathcal{I}_i = \mathcal{I}_i - \langle \mathcal{I} \rangle$ (similarly for $\Delta \mathcal{J}_j$).
Lagrange multipliers $\alpha_\mathcal{I}$ and $\alpha_\mathcal{J}$ were eliminated via the normalization condition.
The MaxEnt distribution of the joint system $P = \{p_{ij}\}$ includes both constraints, so we end with
\begin{equation}\label{eq:3}
\frac{q}{1-q} \ \! [\mathcal{U}_q(P)]^q p_{ij}^{q-1} - \alpha_{\mathcal{IJ}} -  \beta_\mathcal{I} \mathcal{I}_i -  \beta_\mathcal{J} \mathcal{J}_j \ = \ 0\, .
\end{equation}
By inserting~\eqref{eq:1}-\eqref{eq:2} into~\eqref{eq:3}, we obtain
\begin{eqnarray}
&&\mbox{\hspace{-10mm}}[p_{ij} \, \mathcal{U}_q(P)]^{q-1} \ - \ 1 \nonumber \\[1mm]
&&\mbox{\hspace{-5mm}}= \ \left\{[u_{i} \, \mathcal{U}_q(U)]^{q-1} -1\right\} + \left\{[v_{j} \, \mathcal{U}_q(V)]^{q-1}-1\right\}\!,
\end{eqnarray}
which can be rewritten in terms of the $q$-product $x \otimes_q y = [x^{1-q} + y^{1-q} - 1]_+^{1/(1-q)}$ (with $x,y > 0$) as
\begin{equation}\label{eq:comp}
\frac{1}{p_{ij} \, \mathcal{U}_q(P)} \  = \  \frac{1}{u_{i} \, \mathcal{U}_q(U)} \otimes_q \frac{1}{v_{j} \, \mathcal{U}_q(V)}\, .
\end{equation}
When we apply to (\ref{eq:comp}) the $q$-logarithm we obtain
\begin{eqnarray}
\mbox{\hspace{-3mm}}I_q(P) \ominus_q S_q(P) \ = \ [I_q(U) \ominus_q S_q(U)] \ + \  U\leftrightarrow V \, .
\label{eq:comp.a}
\end{eqnarray}
Here $I_q(r_k) = \ln_q ({1}/{r_k})$ is the Tsallis-type Hartley information,
$S_q(R) = \ln_q \mathcal{U}_q(R)$ is the Tsallis entropy and $x \ominus_q y = (x-y)/[1+(1-q)y]$ is the $q$-difference.
Note that~(\ref{eq:comp.a}) represents a $q$-deformed version of the additive entropic rule.
For $q \rightarrow 1$, we recover the relation $p_{ij} = u_i v_j$, which implies the {\em independence} of systems.
To proceed, we now re-express~\eqref{eq:comp} in terms of escort distributions $P_{ij}(q)$, $U_i(q)$ and $V_j(q)$ as
\begin{equation}\label{eq:compes}
\frac{P_{ij}(q)}{p_{ij}} \ = \ \frac{U_{i}(q)}{u_{i}}\ + \ \frac{V_{j}(q)}{v_{j}} \ - \ 1\, .
\end{equation}
The factorization rule $p_{ij} = u_i v_j$ valid in the $q \rightarrow 1$ limit emerges when $\frac{\mathrm{d}}{\mathrm{d} q}\cdots |_{q = 1}$
is taken on both sides of~\eqref{eq:compes}. For $q \neq 1$, we will see that intrinsic correlations are present
even when constraints are independent.
%
%

{\em {Issue of correlations.}}~---~
In order to illustrate the connection to correlations, we investigate the regime where $q$ is close to $1$.
So, we expand a generic escort distribution $R_k(q)$
in the vicinity of  $q=1$ ($q \equiv 1 + \Delta q$), as
\begin{eqnarray}
&&\mbox{\hspace{-9mm}}R_k(q) \ = \ r_k  - r_k \Delta q \ \! [I(r_k) - \Gamma^R_1]^{^{}} \nonumber \\[1mm]
&&\mbox{\hspace{-9mm}}+  r_k \frac{(\Delta q)^2}{2} \ \!\left\{[I(r_k) - \Gamma^R_1]^2 - \Gamma^R_2\right
\} + \mathcal{O}((\Delta q)^3)  ,
\label{eq:escort}
\end{eqnarray}
where $I(r_k) \equiv I_1(r_k)= \ln (1/r_k)$ is the Hartley information
of $k$-th event and $\Gamma_n$ are the cumulants obtained from the generating function $G^R(q) = \ln \sum_k r_k^q$, i.e.,
$\Gamma^R_n = \frac{\mathrm{d}^n G^R(q)}{\mathrm{d}q^n}  |_{q=1}$. Notably, $\Gamma^R_1 = S = -\sum_k r_k \ln r_k$ is
the SE and $\Gamma^R_2 = \sum_k r_k \ln^2 r_k - (\sum_k r_k \ln r_k)^2$ is the varentropy~\cite{schlogl}.
By inserting~\eqref{eq:escort} into~\eqref{eq:compes} we obtain
\begin{eqnarray}\label{eq:q1}
&&\mbox{\hspace{-10mm}}I(p_{ij}) - I(u_{i}) -  I(v_{j}) - \left(\Gamma_1^P-\Gamma_1^U-\Gamma_1^V\right) \nonumber \\[1mm]
&&= \ \frac{\Delta q}{2} \left\{\left[I(p_{ij})-\Gamma_1^P\right]^2 - \left[I(u_{i})-\Gamma_1^U\right]^2  \right. \nonumber \\[1mm]
&&\mbox{\hspace{4mm}}\left. - \left[I(v_{j})-\Gamma_1^V\right]^2  - \left(\Gamma_2^P-\Gamma_2^U-\Gamma_2^V\right) \right\}.
\end{eqnarray}
It is easy to show that for independent systems one has $\Gamma_k^U + \Gamma_k^V = \Gamma_k^{UV}$, where $UV \equiv U \cup V  = \{ u_i v_j \}_{i,j}^{n,m}$ is the ensuing joint distribution. Thus, the differences $(\Gamma_k^P-\Gamma_k^U-\Gamma_k^V)$ quantify the
correlations in the system. This can be seen by considering $p_{ij} = (1+\epsilon_{ij})u_i v_j$, where
$\max_{ij}|\epsilon_{ij}| \ll 1$. In this case, we have
\begin{eqnarray}
\Gamma_1^P &=& \Gamma_1^U \ + \ \Gamma_1^V \ - \ \mbox{$\frac{1}{2}$} \langle \epsilon^2 \rangle_{0} \ + \ \mathcal{O}(\epsilon^3)\, ,\\
\Gamma_2^P &=& \Gamma_2^U \ + \ \Gamma_2^V \ + \ \langle \epsilon \ln^2(UV)\rangle_{0} \ + \ \mathcal{O}(\epsilon^2)\, ,
\end{eqnarray}
where (see also Supplemental Material~\cite{note})
\begin{eqnarray}
\langle \epsilon^2 \rangle_{0} &=& \sum_{ij} \epsilon^2_{ij} u_i v_j\, ,\label{16.aa}\\
\langle \epsilon \ln^2(UV)\rangle_{0} &=&
 \sum_{ij} \epsilon_{ij} \ln^2 (u_i v_j) u_i v_j\, .\label{16.ab}
\end{eqnarray}
The term $\langle \epsilon^2 \rangle_{0}$ represents the strength of the correlations, and is always non-negative.
The case $\langle \epsilon^2 \rangle_0 = 0$ happens only for independent distributions corresponding to $q=1$.
$\Gamma_2^P$ 
represents a specific heat of the system (e.g., $C_p$ in thermal systems)~\cite{beck,ll}.
Expression (\ref{16.ab}) thus represents the difference in specific heats $\Delta C $  with and without correlations $\epsilon_{ij}$.
A connection of the $q$ parameter with $\epsilon_{ij}$ can be established by inserting~\eqref{16.aa}-\eqref{16.ab} into~\eqref{eq:q1},
multiplying the whole equation by $u_i v_j$ and summing over $i$ and $j$.
At the leading order in $\epsilon$ we get
\begin{equation}
q \ = \ 1 \ - \ 2\frac{\langle \epsilon^2 \rangle_{0}}{ \langle \epsilon \ln^2(UV)\rangle_{0}} \ = \  1\ + \  4\frac{\Delta S}{\Delta C} \, .
\label{20aa}
\end{equation}
Systems where the SJ implicit assumption about
the product rule for disjoint systems fails are, e.g., systems where
the number of accessible states $W(N)$ does not grow exponentially with the number of distinguishable subsystems (e.g., particles), i.e.
$W(N) \neq \mu^{N}, \mu >1$  for $N \gg 1$. 
Indeed, in such cases $W(N+M) \neq W(N)W(M)$  and hence the {\em asymptotic equipartition property}
~\cite{jaynes3,thurner-book}: $\lim_{N\rightarrow \infty}\{[S[P_N] + \ln W(N)]/N \} = 0$, $(P_N = \{ p_i\}_{i=1}^N)$,
is not warranted.
However, sub- (or super-) exponential behavior is often found in
strongly correlated systems in quantum mechanics~\cite{canosa,rajagopal1}, high-energy particle 
physics~\cite{wong,zheng,parvan,biro}, or astrophysics~\cite{atsushi,plastino90}. Let us now focus on two  examples.

{\em Examples:}~---~We consider first a generic 2-qubit quantum system (e.g., a bipartite  spin-$\frac{1}{2}$ system). Starting from
un-entangled states $| 11\rangle, | 10\rangle, | 01\rangle, | 00\rangle$ we pass to the Bell basis of maximally entangled orthonormal states $|\Psi^{\pm} \rangle = \frac{1}{\sqrt{2}}(|00\rangle \pm | 11\rangle)$ and $|\Phi^{\pm} \rangle = \frac{1}{\sqrt{2}}(|01 \rangle \pm |10 \rangle)$. Let us examine the situation where the only available constraint is given by a Bell-CHSH observable~\cite{Quant_1,Quant_2} $B= |\Phi^+\rangle \langle\Phi^+|-|\Psi^- \rangle \langle \Psi^-|$ mean value of which yields the (scaled) CHSH Bell inequality~\cite{Quant_1,Quant_3}.
According to MEP we should maximize $S(\rho) = \left[\mbox{Tr}(\rho^q)\right]^{1/(1-q)}$ ($q>0$) subject to constraints $\mbox{Tr}(\rho) = 1$  and $\mbox{Tr}(\rho B) = b$ with $|b|\leq 1$. The corresponding
MEP state, is given by~\cite{note}
\begin{eqnarray}
&&{\mbox{\hspace{-9mm}}}\rho_{\rm{MEP}} \ \! = \ \! Z^{-1}(x,q)\left[\left(|\Phi^- \rangle \langle \Phi^-| + |\Psi^+ \rangle \langle \Psi^+|  \right) + \right.\nonumber \\
&&\left.{\mbox{\hspace{-9mm}}}(1+ x)^{1/(q-1)}|\Phi^+\rangle \langle\Phi^+| + (1- x)^{1/(q-1)}|\Psi^- \rangle \langle \Psi^-|\right]\!,
\end{eqnarray}
where $x = \beta/\alpha$ is the ratio of Lagrange multipliers and $Z(x,q) = 2 + (1+ x)^{1/(q-1)} + (1- x)^{1/(q-1)}$.
We see that  $\rho_{\rm{MEP}}$ is diagonal in the Bell basis. This Bell-diagonal state is not entangled 
if and only if~\cite{Quant_1}
all its eigenvalues are less than or equal to $\frac{1}{2}$.
From concavity of $(1\pm x)^{1/(q-1)}$ for $q \geq 2$ and ensuing Jensen's inequality it is easy to conclude~\cite{note} that all eigenvalues of $\rho_{\rm{MEP}}$ are $\leq 1/2$.
Consequently, for $q\geq 2$ we obtain that $\rho_{\rm{MEP}}$ is not entangled (i.e. is {\em separable}).
Situation for $q<2$ is not conclusive, though {\em inseparability} can be deduced numerically. Fortunately, the case $q=1$ (i.e. SE case) is accessible analytically~\cite{note}. In this case the eigenvalues of $\rho_{\rm{MEP}}$ are: $p_{\Phi^-} = p_{\Psi^+} = \frac{1}{4}\left(1- b^2\right)$,
$p_{\Phi^+} = \frac{1}{4} \left(1- b\right)^2$, $p_{\Psi^-} = \frac{1}{4} \left(1- b\right)^2$.
So, particularly for $b\in (\sqrt{2} -1,1]$ Shannonian MEP clearly predicts entanglement.
However, one can find a non-MEP state~\cite{Quant_1}, namely
\begin{eqnarray}
\mbox{\hspace{-2mm}}\rho  =   b|\Phi^{+} \rangle \langle\Phi^+| +  \mbox{$\frac{1}{2}$}(1 - b) \left(|\Psi^+ \rangle \langle \Psi^+|  +  |\Phi^{-} \rangle \langle\Phi^-| \right)\!,
\end{eqnarray}
which satisfies the MEP constraint and is {\em separable} for $b \leq \frac{1}{2}$.
Hence, Shannonian MEP predicts entanglement even if  for $b\in (\sqrt{2} -1,
\mbox{$\frac{1}{2}$}]$ there is a {\em separable} state that is fully compatible with the constraining data.

Clearly, the correct inference scheme (such as presumed Shannonian MEP)  should not yield an {\em inseparable}
state if there exists (albeit only theoretically) a {\em separable} state compatible with the constraining data, or else
one may get erroneous results (e.g., in quantum communication) by
trying to use the entanglement inferred by the MEP, while in reality, there is
no entanglement present~\cite{Quant_1}.
Note,  that when the  MEP with $\mathcal{U}_q, q\geq2$ is chosen,
one can avoid the
fake entanglement for any $b \leq 1$.
The reason why Shannonian MEP implies spurious (quantum) correlations is
in that analyzed quantum system does not comply with  SSE due to use of the non-local Bell-CHSH observable.
We note that problems with Shannonian MEP should be generically expected in entangled systems as  entanglement does not conform to
SSI because measurement results on (possibly distant)
non-interacting subsystems (giving according to SJ independent constraints) are still correlated.
Situation should be particularly pressing
in strongly-entangled $N$-partite systems because there  $W(N) \propto N^{\rho}, \rho>0$,
cf.~\cite{tsallis,tsallis-3}.

As a second example we consider the transverse momentum  ($p_{_T}$) distributions of hadrons produced
in $pp$ collisions at very high energies (center-of-mass energies $\sim 10^2-10^3$ GeV) as measured in RHIC and LHC experiments.
The term {\em transverse} relates to the direction of colliding protons.
From particle phenomenology it is known that in these cases the effective number of distinguishable states with energy $E$ shows a sub-exponential
growth~\cite{biro2,biro3}, i.e., $W(E)\sim \exp(\langle N \rangle^{\gamma})$ with $0 < \gamma < 1$ and $\langle \cdots \rangle$
taken with respect to an appropriate multiplicity distribution. SSI (and hence Shannon's MEP) is thus not warranted in these cases. In fact,  the single-particle
$p_{_T}$
distributions are best fitted
by the $q$-gaussian distributions (resulting from MEP based on $\mathcal{U}_q$) with $q \in [1.05, 1.10]$ depending
on the type of the collision~\cite{biro_II,PHENIX,STAR,ALICE_I,ALICE_II}.
In these cases the constraint (\ref{constraint.2}) is represented by the mean of the transverse energy $E_{_T} = \sqrt{{{p}}_{_T}^2 + m^2}$  ($m$ is  hadron's rest mass). Typical picture is that
out of many hadrons produced in a given event only one is selected (system $A$). Remaining $(N-1)$ particles ($N$ is event dependent) act as a kind of a heat bath (HB) (system $B$) described by some {\em apparent}  temperature. In this HB the single-hadron ${{p}}_{_T}$ is effectively distributed according to the Maxwell--J\"{u}ttner  distribution. The final distribution $u(p_{_T})$ is obtained by averaging over many events with distinct
apparent temperatures. Systems $A$ and $B$ are clearly disjoint, but due to event-to-event temperature fluctuations the joint distribution $p(p_{_T}, p_{_B}) \neq u(p_{_T}) v(p_{_B})$, so SSI is indeed violated.
Now, since $q$ is close to $1$, we can consider only the leading order of $\epsilon_{ij}$ in  $(q-1)$, i.e.
$\epsilon_{ij} = (1-q) \beta^2 \Delta E_{i}^{u} \Delta E_{j}^{v}$. From (\ref{20aa}) then follows that~\cite{note}
\begin{eqnarray}
q \ = \ 1 + \frac{\langle N \rangle -1}{\beta^2\langle (\Delta E^{v})^2 \rangle_0 } \ = \ 1 +  \frac{\langle N \rangle -1}{C_V^v}\, .
\label{23.aa}
\end{eqnarray}
where $\langle (\Delta E^{v})^2 \rangle_0 = \partial^2 \log Z^{v} /\partial \beta^2 = C_V^{v}/\beta^2$
[$Z^{v}$ and $C_V^v$ represent partition  function  (i.e., $\mathcal{U}_q(V)$) and heat capacity of the HB] and $\langle N -1 \rangle = \beta \langle E^v \rangle_0$ is the virial relation where  $1/\beta$ is
the kinetic temperature of the hadronic HB.
Note that system $A$ factored  out.  Relations of the type (\ref{23.aa}) frequently appear in phenomenological studies on high-energy $pp$ collisions~\cite{biro_II,wilk,wilkII}.

{\em {Conclusions.}}~---~In summary, we have shown that the SJ axiomatization of the inference
rule does account for substantially wider class of entropic functionals than just SE. The root cause could be retraced
to unreasonably strong assumptions employed by SJ in their proof --- assumptions that go beyond the original SJ axioms.
In particular we have shown that Shannonian MEP is singled out as an unique method of statistical inference only insofar as an extra axiom of \emph{strong system independence} is added to the SJ desiderata.
While, for systems where state-space scales exponentially with its size
(as, e.g., in (quasi-) ergodic systems) 
SE is the only entropy compatible with SJ axioms,
for systems with sub- (super-) exponential growth the assumption of SSI is not
justified and the original proof of SJ needs revision. 
In our revised version of the proof we identified a one-parameter class of admissible entropies whose utility was illustrated with
two phenomenologically relevant examples; $2$-qubit quantum system and hadron productions in $pp$ collisions.

P.J. and J.K. were supported by the Czech Science Foundation (GA\v{C}R), Grant 17-33812L.
J.K. was also supported by the Austrian Science Foundation (FWF) under project I3073.

\end{document}



\title{Supplemental Material for ``Maximum Entropy Principle in statistical inference:\\ case for non-Shannonian entropies''}

\author{Petr Jizba}
\email{p.jizba@fjfi.cvut.cz}
\affiliation{FNSPE,
Czech Technical University in Prague, B\v{r}ehov\'{a} 7, 115 19, Prague, Czech Republic}

\author{Jan Korbel}
\email{jan.korbel@meduniwien.ac.at}
\affiliation{Section for Science of Complex Systems, Medical University of Vienna, Spitalgasse 23, 1090 Vienna, Austria}
\affiliation{Complexity Science Hub Vienna, Josefst\"{a}dter Strasse 39, 1080 Vienna, Austria}
\affiliation{FNSPE,
Czech Technical University in Prague, B\v{r}ehov\'{a} 7, 115 19, Prague, Czech Republic}


\vspace{2mm}
\maketitle

{\bf Note:} equations and citations that are related to the main text are shown in red.

\section*{Proof of an inference rule based on SJ axioms; part I --- the sum form {\cdred (3)}}

For the sake of completeness, we demonstrate here that {\em axiom of uniqueness}, {\em permutation invariance}
and {\em subset invariance} lead to the sum form of the entropy {\cdred (3)}, i.e.
\begin{equation}
\mathcal{U}(P) \ = \ \sum_{i=1}^{n} g(p_i) \ \sim  \ f\left(\sum_{i=1}^{n} g(p_i)\right) .
\label{SM.1.a}
\end{equation}
where $f$ is a monotonic function. In our exposition we will loosely follow the main steps from the SJ
original proof~{\cite{shore}}. Let us first remind several observations. First,
from the {\em permutation invariance} axiom we obtain that $\mathcal{U}(P)$ is a symmetric
function of $P = \{p_1,\dots,p_n\}$. Let us now focus on the third axiom, i.e., {\em subset invariance}.
Let us choose a subset $M \subseteq N = \{1,\dots,n\}$. We further consider a constraint on the subset $M$, i.e.,
\begin{equation}
\sum_{j \in M} \mathcal{I}^M_{j} p_j \ = \ \langle \mathcal{I}^M \rangle \ \equiv \ {I}^M\, .
\end{equation}
This constraint can be easily expanded to the whole state-space, when we define $\mathcal{I}^M_j = 0$
for $j \notin M$. Conditional probabilities on $M$ are defined as
\begin{equation}
q^{M}_j \ = \ P(x_j|x_j \in M) \ = \ \frac{p_j}{\sum_{j \in M} p_j}\, .
\end{equation}
Let us denote $r = \sum_{j \in M} p_j$. The conditional distribution on the set $N\smallsetminus M$ is then defined as $q_j^{N\smallsetminus M} = {p_j}/{(1-r)}$.
From the {\em subset independence} axiom, it should not matter whether we use MEP for the conditional distributions
$Q^M = \{q^M_1,\dots,q^M_m\}$ and $Q^{N\smallsetminus M} = \{q^{N\smallsetminus M}_1,\dots,q^{N\smallsetminus M}_m\}$ or
for the full distribution $P$. The conditional distribution $Q^M$ does not depend on $Q^{N\smallsetminus  M}$ neither on $n$,
because it is a solution of the MEP on the reduced space $M$.

Let us start by examining the MEP in terms of above $Q$s. From the {\em subset independence} axiom,
we obtain that
the MaxEnt procedure is equivalent to maximization of two functionals
\begin{eqnarray}
&&\mathcal{U}(Q^M) \ - \ \alpha_M \sum_{j \in M} q^{M}_j \ - \ \beta_M \sum_{j \in M} r \mathcal{I}^M_j q^{M}_j\, ,\\
&&\mathcal{U}(Q^{N\smallsetminus M}) \ - \ \alpha_{N \smallsetminus M} \sum_{j \in N \smallsetminus M} q^{M}_j\, ,
\end{eqnarray}
which yield equations
\begin{eqnarray}
&&\frac{\partial \mathcal{U}(Q^M)}{\partial q^M_i} \ = \ \alpha_M(Q^{M}) + r \beta(Q^{M}) \, \mathcal{I}^M_i \ = \ f(q^M_i,Q^{M-i}) \ \equiv \ f^{M}_i\, ,\\
&&\frac{\partial \mathcal{U}(Q^{N\smallsetminus M})}{\partial q^{N\smallsetminus M}_j} \ = \ \alpha_M(Q^{N\smallsetminus M})\, ,
\end{eqnarray}
where $Q^{M-i}$ is  $Q^{M}$ without $q^{M}_i$. The form of $f$ is a consequence of permutation invariance of $\mathcal{U}$.

Alternatively, we can formulate the problem for the full distribution as $\mathcal{U}(P) - \alpha \sum_{j} p_ -  \sum_{j} \mathcal{I}^M_{j} p_j$ leading to
\begin{equation}
\frac{\partial \mathcal{U}(P)}{\partial p_i} \ = \ \alpha(P) \ + \  \beta(P) \, \mathcal{I}^M_i \ = \ h(p_i,P^{N-i}) \ \equiv \ h_i\, ,
\end{equation}
where $P^{N-i}$ is $P$ without $p_i$. Let us now take $i \in M$ and express ${\partial \mathcal{U}(P)}/{\partial p_i}$ in terms of $Q^M$, i.e.
\begin{equation}
\frac{\partial \mathcal{U}(P)}{\partial p_i}\ = \ \sum_{j \in M} \frac{\partial \mathcal{U}(Q^M)}{\partial q^M_j} \ \! \frac{\partial q^M_j(P)}{\partial p_i}\, .
\end{equation}
In fact, since
\begin{eqnarray}
\frac{\partial q^M_i(P)}{\partial p_i} \ = \ \frac{1}{r} \ - \  \frac{p_i}{r^2}\, ,\;\;\;\;\;\;\; \mbox{and} \;\;\;\;\;\;\;
\frac{\partial q^M_j(P)}{\partial p_i}  \ = \ -  \frac{p_j}{r^2}\;\;\;
\;\;\;\;{\mbox{for}} \;\;\;\; j \neq i\, ,
\end{eqnarray}
we have
\begin{equation}
h_i \ = \ \frac{f^M_i}{r} \ - \ \frac{1}{r}\sum_{j \in M} q_j^{M} f^M_j\, .
\label{SM.12.a}
\end{equation}
To proceed, we define the quantity
\begin{equation}
W_{ijk} \ = \ \frac{h_i-h_j}{h_k - h_j} \ = \ \frac{f^M_i-f^M_j}{f^M_k-f^M_j}\, ,
\label{SM.13.a}
\end{equation}
for $i,j,k \in M$.
It is easy to check that $W$ fulfills the following functional relation
\begin{equation}
W_{ijk} \ = \ \frac{W_{irv}-W_{jrv}}{W_{krv}-W_{jrv}}\, .
\end{equation}
This is the so-called Cantor functional equation which has the general solution~\cite{cantor}
\begin{equation}
W_{ijk} \ = \ \frac{u(p_i) - u(p_j)}{u(p_k)-u(p_j)}\, ,
\label{SM.15.a}
\end{equation}
where $u(x)$ is an arbitrary function. By employing (\ref{SM.13.a}) and (\ref{SM.15.a}) we can deduce that
\begin{equation}
h_i \ = \ \frac{\partial \mathcal{U}(P)}{\partial p_i} \ = \ s(P) u(p_i) \ + \ z(P)\, .
\label{SM.16.a}
\end{equation}
Here $s(P) = ({h_k-h_j})/[{u(p_k) - u(p_j)}]$ and $z(P)= h_j - u(p_j) s(P)$ ($j \in M$ is arbitrary).
Note that $s(P)$ and $z(P)$
do not depend on an actual choice of indices $k,j \in M$. Indeed, for $s(P)$ we may write
\begin{eqnarray}
s(P) \ = \ \frac{h_k - h_j}{u(p_k) - u(p_j)} \ = \  \frac{h_l - h_j}{u(p_l) - u(p_j)} \ = \ \frac{h_j - h_l}{u(p_j) - u(p_l)} \ = \  \frac{h_r - h_l}{u(p_r) - u(p_l)}\, ,
\end{eqnarray}
where 2nd and 4th equality results from the combination of Eqs.~(\ref{SM.13.a}) and (\ref{SM.15.a}) while in 3rd equality
we have simultaneously multiplied numerator and denominator by $-1$. Indices $l,r \in M$ are arbitrary. Independence of $z(P)$ on $j$ follows then immediately from the
invariance of $h_i$ under arbitrary permutation of elements  $p_j \in P^{N-i}$.


Since the above analysis does not depend on the cardinality of $M$, we can set now $M=N$.
From (\ref{SM.16.a}) thus follows
%
\begin{equation}
{\mathrm{d}\, \mathcal{U}(P)} \ = \ \sum_i h_i {\mathrm{d}\,  p_i}\ = \ s(P) \sum_i u(p_i) {\mathrm{d}\, p_i}\ + \ z(P)\ \!{\mathrm{d}}\!\left(\sum_i p_i\right) \ = \ s(P) {\mathrm{d} G(P)}\, ,
\label{SM.17.a}
\end{equation}
where ${\mathrm{d} \, \mathcal{U}(P)}$ is the exact differential and $G(P) = \sum_i g(p_i)$ with $g(x)$ being an antiderivative of $u(x)$, i.e., $u(x) = {\mathrm{d} g(x)}/{\mathrm{d} x}$.
Result (\ref{SM.17.a}) directly shows that $\mathcal{U}(P)$ must be a function of $G(P)$,
and the function itself, say $f$, is obtained by solving the equation ${s(P)} = {\mathrm{d}}f(G)/{\mathrm{d}}G$. Moreover, the $f$-function must be monotonic,
because $s(P)$ does not flip the sign. Indeed,
if $P$ and $Q$ would be two distributions such that $s(P) >0$ and $s(Q) < 0$ then for any trajectory $P(t)$ such that $P(0) = P$ and $P(1) = Q$
would need to exist a point $t_0 \in (0,1)$, such that  $s(P(t_0)) = 0$. However, from the definition of $s(P)$ it follows that $s(P) = 0$ if and only if $h_k = h_j$ for all $i,j$,
which happens only for a single point in the probability simplex, namely the point corresponding to a uniform distribution $P_u$. Since the general trajectory connecting any two
points in the probability simplex does not cross point $P_u$, $s(P)$ must always have the same sign with $P_u$ being merely extremal point of $s(P)$.
This concludes the proof.

\section*{Proof of an inference rule based on SJ axioms; part II --- axiom 4 without SJ factorization assumption}


Axiom 4 can be restated as follows: Let us consider two systems $A$ and $B$ with
the elementary outcomes  $\{{a}_i\}_{i=1}^n$ and $\{{b}_j\}_{j=1}^m$, respectively.
We denote the joint distribution of the composed system $A \cup B$ as $ P(A = a_i, B = b_j) = p_{ij}$ and corresponding marginal
distributions as $u_i = \sum_j p_{ij}$ and  $v_j = \sum_i p_{ij}$. We further consider two {\em independent} constraints,
affiliated with the two subsystem, i.e.,  $\sum_{i=1}^n  \mathcal{I}_i u_i \equiv {I}$ and $\sum_{j=1}^m \mathcal{J}_j v_j \equiv {J}$.
Each constraint can be naturally rewritten in terms of the joint distribution $p_{ij}$, so that the Lagrange functional for the full
distribution can be expressed (modulo equivalency condition) as
\begin{equation}\label{eq:lag}
\sum_{ij} g(p_{ij}) - \alpha \sum_{ij} p_{ij} - \beta_\mathcal{I} \sum_{ij} \mathcal{I}_i p_{ij} - \beta_\mathcal{J} \sum_{ij} \mathcal{J}_j p_{ij}\, ,
\end{equation}
which leads to
\begin{equation}\label{eq:a}
g'(p_{ij}) - \alpha - \beta_\mathcal{I} \mathcal{I}_i - \beta_\mathcal{J} \mathcal{J}_j \ = \ 0\, .
\end{equation}
Similarly, the same problem can be formulated directly in terms marginal distributions $u_i$ and $v_j$.
This immediately implies that the entropy functional is {\em composable}, i.e.,
$\mathcal{U}(p_{ij}) = \Phi(\mathcal{U}(u_i),\mathcal{U}(v_j))$ for some
function $\Phi$. By employing the assumption of the axiom 4, i.e., that independent systems
with related constraints considered separately (i.e., in terms of
marginal distributions) should be equivalent to (\ref{eq:lag}), we see that
$\Phi$ must be {\em proportional} to the multiplication operation, i.e.
\begin{equation}
\mathcal{U}(p_{ij}) \ = \ \mathcal{U}(u_i)\, \mathcal{U}(v_j) \ = \ \sum_i g(u_i) \sum_j g(v_j)\, .
\label{eq:6b}
\end{equation}
As a result, we obtain that the Lagrange functional is equal to
\begin{eqnarray}
\sum_{ij} g(u_i)g(v_j) \ - \ \alpha \sum_{ij} p_{ij}
\ - \ \beta_\mathcal{I} \sum_{i} \mathcal{I}_i u_i \ - \ \beta_\mathcal{J} \sum_{j} \mathcal{J}_j v_j\, .
\label{SM.22}
\end{eqnarray}
Our aim is now to rewrite the Lagrange functional in terms of $u_i$ and $v_j$. We first note that the term corresponding
to the normalization condition can be rewritten as $\alpha \sum_{ij} u_i v_j$ because $\sum_{ij} p_{ij} = \sum_i u_i = \sum_jv_j = \sum_{ij} u_iv_j$.
Note that this is true without assuming that $p_{ij} = u_iv_j$. In other words, we do not need to assume anything about the
actual factorization rule. This is indeed the crux of this work.
Second, from {\em permutation invariance} of entropy and the fact that the subsystems are disjoint, we obtain that the Lagrange functional is symmetric under permutation of states. This is easily seen when (\ref{SM.22}) is rewritten as:
\begin{eqnarray}
\sum_{ij} g(u_i)g(v_j) \ - \ \alpha \sum_{ij} u_i v_j
\ - \ \beta_\mathcal{I} \sum_{ij} \mathcal{I}_i u_i v_j \ - \ \beta_\mathcal{J} \sum_{ij} \mathcal{J}_j u_i v_j\, .
\end{eqnarray}
%
%
The extremization procedure provides the equations determined by derivatives under $u_i$ and $v_j$, respectively, i.e.

\begin{eqnarray}
g'(u_i) \sum_{j}g(v_j) \ - \ \alpha \sum_{j} v_j
\ - \ \beta_\mathcal{I} \mathcal{I}_i \sum_{j} v_j \ - \ \beta_\mathcal{J} \sum_{j} \mathcal{J}_j v_j\ = \ 0\, ,\\
g'(v_j) \sum_{i} g(u_i) \ - \ \alpha \sum_{i} u_i
\ - \ \beta_\mathcal{I} \sum_{i} \mathcal{I}_i u_i \ - \ \mathcal{J}_j \beta_\mathcal{J} \sum_{i}  u_i\ = \ 0\,  .
\end{eqnarray}
By taking another derivative w.r.t. $v_j$ and $u_i$, respectively, we obtain a single identity
\begin{equation}\label{eq:b}
g'(u_{i}) g'(v_{j}) \ - \ \alpha \ - \ \beta_\mathcal{I} \mathcal{I}_i \ - \ \beta_\mathcal{J} \mathcal{J}_j \ = \ 0\, .
\end{equation}
By comparing~\eqref{eq:a} with \eqref{eq:b} we obtain the multiplicative Cauchy functional equation $g'(u_i v_j) = g'(u_i) g'(v_j)$, which has the general solution $g'(x) = x^r$ for any $r\in \mathbb{R}$. Consequently,
$g(x) = x^{r+1}/(r+1) + a$ with $a \in \mathbb{R}$. So, any entropy functional consistent with
SJ axioms should be {\em equivalent} to $\sum_i p_i^q$, where $q=r+1$.

At this stage one should distinguish two cases, i.e. $q \leq 1$, and $q \geq 1$. For $q \leq 1$, the function $g(x)=x^q$ is concave, while for $q \geq 1$ it is convex. Thus, one might merge both cases into a single
functional of the form (cf. also~\cite{uffink})
%
\begin{equation}
\mathcal{U}_q(P) \ = \ \left(\sum_i p_i^q\right)^{1/(1-q)}\, ,
\end{equation}
which is Schur-concave for $q > 0$. It should be stressed that the Schur-concavity is a sufficient condition for {\em maximality} axiom
(see, e.g., Refs.~\cite{marshal,roberts}). The case $q > 0$, is ruled out by the {\em maximality axiom}, because for $q\leq 0$ we obtain a local minimum for $P_u$, instead of the required
maximum.
This closes the proof.
In passing we might note that for $q \rightarrow 1$ we obtain $\mathcal{U}_1(P) = \exp[\mathcal{H}(P)]$ where $\mathcal{H}(P) = - \sum_i p_i \log p_i$ is the Shannon entropy.
Functional $\mathcal{U}_1(P)$ is knows as Shannon's entropy power. \\[2mm]

Let us finally stress that should we have assumed (in addition to the SJ axioms) the factorization rule $p_{ij} = u_iv_j$,
then as shown in the main text {only} $\mathcal{U}_1(P)$ would be consistent with this requirement. Then, Shannon's entropy
(modulo equivalency condition) would be indeed the only consistent candidate for MEP.
So, the uniqueness of Shannon's entropy can be ensured {only} when an extra axiom is added to SJ consistency axioms, namely:  \\[2mm]

\emph{Strong system independence:} Whenever two subsystems of a system are disjoint, we can treat the subsystems in terms of independent distributions.

\section*{Expansion of cumulants for weak correlations}

Here we show how one can obtain equations {\cdred (21)} and {\cdred (22)} in the main text.
Let us remind the definition of the cumulants $\Gamma_1^R$ and $\Gamma_2^R$ (the varentropy), namely
\begin{eqnarray}
\Gamma_1^R \ = \ - \sum_k r_k \ln r_k\, , \;\;\;\;\;\;\;
\Gamma_2^R \ = \ \sum_k r_k \ln^2 r_k \ - \ \left(\sum_k r_k \ln r_k\right)^2\! .
\end{eqnarray}
We further consider two subsystems with respective distributions $U=\{u_i\}_i^n$ and $V = \{v_j\}_j^m$. The resulting compound system is described
by the probability distribution $P = \{p_{ij}\}_{i,j}^{n,m}$. Let us now assume that the systems are {\em weakly correlated} so that we
can write
\begin{equation}
p_{ij}\ = \ (1 \ + \ \epsilon_{ij})u_i v_j\, ,
\label{eq:cor}
\end{equation}
for some matrix ${\epsilon}_{ij}$ with the {\em{max norm}} satisfying
$|\!|{\epsilon}|\!|_{\rm{max}} \ = \ \max_{ij} |\epsilon_{ij}| \ll 1$. The corresponding $\Gamma_1^P$  reads
\begin{eqnarray}
\Gamma_1^P
&\approx& - \sum_{ij} (1 \ + \ \epsilon_{ij})u_i v_j \ \! \left[\ln (u_i v_j) \ + \ \epsilon_{ij} \ - \ \mbox{$\frac{1}{2}$}\epsilon_{ij}^2 \right] \nonumber \\
&=& -\sum_{ij} u_i v_j \ln(u_i v_j) - \sum_{ij} \epsilon_{ij} u_i v_j \ln (u_i v_j) - \sum_{ij} \epsilon_{ij}u_i v_j - \mbox{$\frac{1}{2}$} \sum_{ij} \epsilon_{ij}^2 u_i v_j \, ,
\label{III.29.a}
\end{eqnarray}
where we used that $\ln (1+\epsilon_{ij}) \approx \epsilon_{ij}$. Clearly, only the first and last term are non-zero.
This can be easily seen by realizing that Eq.~\eqref{eq:cor} implies
$\sum_i \epsilon_{ij} u_i  = \sum_j \epsilon_{ij} v_j  =  0$ and
hence for any two non-singular functions $f$ and $g$  we have
\begin{equation}
\sum_{ij} \epsilon_{ij}u_i v_j [f(u_i)+g(v_j)] \  = \ \sum_i f(u_i) u_i \sum_j \epsilon_{ij} v_j \ + \ \sum_j g(v_j) v_j \sum_i \epsilon_{ij} u_i \ = \ 0\, .
\end{equation}
The second and third term in Eq.~\eqref{III.29.a} have exactly this form and are therefore zero. The first term can be rewritten in terms of cumulants
of the marginal distributions and the last term is just the average value of $\epsilon^2_{ij}$. By using the fact that $v_j$ and $u_i$
are positive, then the last term is zero if and only if $\epsilon_{ij}=0$ for all $i$, $j$, i.e., for {\em independent} systems.
Consequently, one can rewrite $\Gamma_1^P$ as
\begin{equation}
\Gamma_1^P \ = \  \Gamma_1^U \ + \ \Gamma_1^V \ - \ \mbox{$\frac{1}{2}$} \langle \epsilon^2 \rangle_0 \ + \ \mathcal{O}(\epsilon^3)\, ,
\end{equation}
where $\langle \epsilon^2 \rangle_{0} = \sum_{ij} \epsilon^2_{ij} u_i v_j$.
Along the same lines we can write also $\Gamma_2^P$, namely
\begin{eqnarray}
\Gamma_2^P &=& \sum_{ij} p_{ij} \ln^2 p_{ij} - [\Gamma_1^P]^2
\ \approx \ \sum_{ij}(1+\epsilon_{ij})u_i v_j  \left[\ln^2(u_i v_j)+ 2 \epsilon_{ij} \ln(u_i v_j) + \epsilon_{ij}^2\right] - \left(\Gamma_1^U + \Gamma_1^V - \mbox{$\frac{1}{2}$}\langle \epsilon^2 \rangle_0 \right)^{\!2}\nonumber \\
&=& \sum_{ij} u_i v_j \left(\ln^2 u_i + 2 \ln u_i \ln v_j + \ln^2 v_j\right) + \sum_{ij} \epsilon_{ij} u_i v_j \ln^2(u_i v_j) - \left(\Gamma_1^U\right)^{\!2} - \left(\Gamma_1^V\right)^{\!2} - 2 \Gamma_1^U\Gamma_1^V + \mathcal{O}(\epsilon^2)\, ,
\end{eqnarray}
where we neglected all terms containing $\epsilon_{ij}^2$ and higher orders. Putting everything together, we obtain
\begin{equation}
\Gamma_2^P \ = \ \Gamma_2^U \ + \ \Gamma_2^V \ + \  \langle \epsilon \ln^2(UV)\rangle_{0} \ + \ \mathcal{O}(\epsilon^2)\, ,
\end{equation}
where $\langle \epsilon \ln^2(UV)\rangle_{0} = \sum_{ij} \epsilon_{ij} \ln^2 (u_i v_j) u_i v_j = 2 \langle \epsilon \ln(U)\ln(V)\rangle_{0}$.

\section*{Calculation of $\rho_{\rm{MEP}}$}

Here we calculate $\rho_{\rm{MEP}}$. The variation of the related Lagrange functional yields the equation
\begin{eqnarray}
0 \ = \ \frac{q}{1-q} \ \![\mathcal{U}_q(\rho)]^{q} \ \! \mbox{Tr}\left(\rho^{q-1} \delta \rho\right) - \alpha  \mbox{Tr}\left(\delta \rho\right) - \beta \mbox{Tr}\left(B \delta \rho\right)\, ,
\end{eqnarray}
which gives [cf. {\cdred (5)} and {\cdred (6)}]
\begin{eqnarray}
\frac{q}{1-q} [\mathcal{U}_q(\rho)]^{q} \ \! \rho^{q-1} -\alpha - \beta B \ = \ 0\, ,
\label{35ac}
\end{eqnarray}
and hence
\begin{eqnarray}
\rho_{\rm{MEP}} \ = \  \left\{\frac{(1-q)}{q}\ \![\mathcal{U}_q(\rho_{\rm{MEP}})]^{-q}\left[\alpha + \beta B   \right]\right\}^{1/(q-1)} \ = \ \frac{\left(\alpha + \beta B
\right)^{1/(q-1)}}{\mbox{Tr}\left[\left(\alpha + \beta B   \right)^{1/(q-1)}\right]}\, .
\label{36abc}
\end{eqnarray}
Let us note that when we multiply (\ref{35ac}) by $\rho$ and simultaneously take $\mbox{Tr} \cdots$ we obtain
\begin{eqnarray}
\alpha \ = \ \frac{1}{1-q}  \mathcal{U}_q(\rho_{\rm{MEP}}) \ - \ \beta b \, ,
\end{eqnarray}
where $b = \mbox{Tr}(B \rho)$. This directly provides the forms {\cdred{(7)}} and {\cdred{(8)}}.
%
%

To proceed we notice that the Bell-CHSH observable $B$ is a (scaled) {\em involutory} operator, namely $B^{2n} \equiv
A = |\Phi^+\rangle \langle\Phi^+|+|\Psi^- \rangle \langle \Psi^-|$ and $B^{2n-1} = B$ with $n = 1,2, \ldots$. By employing the binomial expansion we get
\begin{eqnarray}
\mbox{\hspace{-5mm}}(\alpha + \beta B )^{1/(q-1)} &=& \alpha^{1/(q-1)} \sum_{k=0}^{\infty} \left(
                                                                          \begin{array}{c}
                                                                            \frac{1}{q-1} \\
                                                                            k \\
                                                                          \end{array}
                                                                        \right) \left(\frac{\beta}{\alpha}\right)^k B^k \nonumber \\[2mm]
                                                                        &=&
\alpha^{1/(q-1)} \left(|\Phi^-\rangle \langle\Phi^-|+|\Psi^+ \rangle \langle \Psi^+|  \right)
 + (\alpha + \beta)^{1/(q-1)} |\Phi^+\rangle \langle\Phi^+| +  (\alpha - \beta)^{1/(q-1)}  |\Psi^- \rangle \langle \Psi^-| ,
\end{eqnarray}
where
\begin{eqnarray}
\left(
                                                                          \begin{array}{c}
                                                                            \frac{1}{q-1} \\
                                                                            k \\
                                                                          \end{array} \right) \ = \ \frac{\frac{1}{q-1}\left(\frac{1}{q-1}-1\right)\left(\frac{1}{q-1}-2\right)
                                                                          \cdots\left(\frac{1}{q-1}-k + 1\right)}{k!}\, ,
\end{eqnarray}
is the (generalized) binomial coefficient. With this we can write
\begin{eqnarray}
\rho_{\rm{MEP}} &=&  \frac{\alpha^{1/(q-1)} \left(|\Phi^-\rangle \langle\Phi^-|+|\Psi^+ \rangle \langle \Psi^+|  \right) + \ (\alpha + \beta)^{1/(q-1)} |\Phi^+\rangle \langle\Phi^+|
\ + \ (\alpha - \beta)^{1/(q-1)}  |\Psi^- \rangle \langle \Psi^-|}{2 \alpha^{1/(q-1)}  + (\alpha + \beta)^{1/(q-1)} + (\alpha - \beta)^{1/(q-1)}}\nonumber \\[3mm]
&\equiv&  p_{\Phi^-} |\Phi^-\rangle \langle\Phi^-| \ + \ p_{\Psi^+}|\Psi^+ \rangle \langle \Psi^+|  \ + \ p_{\Phi^+} |\Phi^+\rangle \langle\Phi^+|  \ + \ p_{\Psi^-} |\Psi^- \rangle \langle \Psi^-|\, .
\label{41.abc}
\end{eqnarray}
Consequently, $\rho_{\rm{MEP}}$ is diagonal in the Bell basis. Note that the {\em uniqueness theorem} requires that $\alpha \geq |\beta|$, which also ensures that $\rho_{\rm{MEP}}$
has real and positive spectrum. From (\ref{41.abc}) we see that $p_1$ and $p_2$ are clearly $\leq 1/2$.
In addition, for $q \geq 2$ we can conclude that $p_3$ and $p_4$ are also $\leq 1/2$. Indeed, from concavity of $x^{1/(q-1)}$ for $q \geq 2$ we have
\begin{eqnarray}
\alpha^{1/(q-1)} \ = \ \left[\mbox{$\frac{1}{2}$}(\alpha + \beta) + \mbox{$\frac{1}{2}$}(\alpha - \beta)\right]^{1/(q-1)} \ \geq \ \mbox{$\frac{1}{2}$}(\alpha +
\beta)^{1/(q-1)} + \mbox{$\frac{1}{2}$}(\alpha - \beta)^{1/(q-1)}\, ,
\end{eqnarray}
where the last inequality is due to Jensen's inequality for concave functions. Hence
\begin{eqnarray}
p_{\Phi^+} \ = \ \frac{(\alpha + \beta)^{1/(q-1)}}{2 \alpha^{1/(q-1)}  + (\alpha + \beta)^{1/(q-1)} + (\alpha - \beta)^{1/(q-1)}} \ \leq \ \frac{(\alpha + \beta)^{1/(q-1)}}{2 (\alpha + \beta)^{1/(q-1)} + 2
(\alpha - \beta)^{1/(q-1)}} \ \leq \  \frac{1}{2}\, ,
\end{eqnarray}
and similarly for $p_{\Psi^-}$.  Consequently, according to {\cdred [65]} we have
for $q\geq 2$ that $\rho_{\rm{MEP}}$ is not entangled (i.e. is {\em separable}).

Situation for $q<2$ is not conclusive and we should determine the Lagrange multipliers to be more specific. To this end we notice
that in $\rho_{\rm{MEP}}$ it is only the ration $\beta/\alpha$ that needs to be determined. By employing the constraint $b = \mbox{Tr}(B\rho)$
we obtain from (\ref{41.abc})
%
\begin{eqnarray}
b \ = \ \frac{\left(1 + \frac{\beta}{\alpha}\right)^{1/(q-1)} \ -
\  \left(1 - \frac{\beta}{\alpha}\right)^{1/(q-1)}}{2 \ + \ \left(1 + \frac{\beta}{\alpha}\right)^{1/(q-1)} \ + \ \left(1 - \frac{\beta}{\alpha}\right)^{1/(q-1)}}\, .
\end{eqnarray}
This is a transcendental equation for $\beta/\alpha$  which allows, at least numerically, to solve $\rho_{\rm{MEP}}$ for desired values of $q$ and $b$.
Analytical form of the solution for generic $q$ and $b$ is not possible  but fortunately the case $q=1$ (i.e. Shannon's case) can be done quite easily. In this case we obtain
\begin{eqnarray}
p_{\Phi^-}  \ = \ p_{\Psi^+} \ = \  \mbox{$\frac{1}{4}$} \left(1- b^2\right)\, ,\;\;\;\;  p_{\Phi^+} \ = \  \mbox{$\frac{1}{4}$} \left(1- b\right)^2\, ,\;\;\;\; p_{\Psi^-} \ = \  \mbox{$\frac{1}{4}$} \left(1+ b\right)^2\, .
\end{eqnarray}
So, particularly for $b > \sqrt{2} -1$ we have entanglement (i.e. the MEP state is {\em inseparable}) or, in other words, for $b\in (\sqrt{2} -1,1]$ Shannonian MEP predicts entanglement even if there exists for $b\in (\sqrt{2} -1,
\mbox{$\frac{1}{2}$}]$ a {\em separable} state
that is fully compatible with the constraining data.